# Decrypting SSL/TLS Traffic for Hidden Threats Detection

Tamara Radivilova [1], Lyudmyla Kirichenko [2], Dmytro Ageyev [3], Maxim Tawalbeh [4], Vitalii Bulakh [5]
[1,2,3,4,5] Kharkiv National University of Radioelectronics, 14, ave.Nauki, Kharkiv, Ukraine,
tamara.radivilova@gmail.com, lyudmyla.kirichenko@nure.ua, dmytro.aheiev@nure.ua, tavalbeh@icloud.com
https://orcid.org/0000-0001-5975-0269, http://orcid.org/0000-0002-2426-0777, https://orcid.org/0000-0002-2686-3854

*Abstract*—The paper presents an analysis of the main mechanisms of decryption of SSL/TLS traffic. Methods and technologies for detecting malicious activity in encrypted traffic that are used by leading companies are also considered. Also, the approach for intercepting and decrypting traffic transmitted over SSL/TLS is developed, tested and proposed. The developed approach has been automated and can be used for remote listening of the network, which will allow to decrypt transmitted data in a mode close to real time.

*Keywords—SSL/TLS protocol, vulnerability, attacks, network intrusion detection system, decryption approach*

## I. Introduction

Currently there is a trend of increasing the proportion of encrypted traffic. Now, according to Cisco, 60% of Internet traffic is encrypted, and according to Gartner forecasts, by 2019, 80% of the traffic will be [1, 2]. It is necessary to ensure the privacy of citizens, to keep information in secret, to comply with the requirements of the law. But attackers also use encryption to bypass the mechanisms for detecting their unauthorized activity, hiding interaction with the command servers of malicious programs and for other tasks. [1-3]. Skilled criminals can steal data inside encrypted traffic, successfully exploit vulnerabilities, manage command and control communications avoiding detection. With such an increase in attacks in encrypted traffic, it is necessary to create an approach (method) that can decrypt and verify encrypted traffic. This will help to neutralize the danger of hidden threats.

Security products can not see what happens in encrypted traffic (according to the Ponemon Institute, 64% of companies can not detect malicious code in encrypted traffic). To penetrate into encrypted connections, organizations often use the Man-in-the-Middle attack for legal purposes, but this is not legal (violation of the privacy of correspondence or other legislation requirements to ensure the confidentiality of information). [4]

Usually, a gateway or cluster of gateways is installed on the perimeter of a corporate or departmental network. They perform interception and decryption of data within the company's network to protect against attacks using SSL/TLS (SSL - Secure Sockets Layer, TLS - Transport Layer Security) for the transmission of malicious content, and for analyzing transmitted data by Intrusion Detection System (IDS) (according to the Ponemon Institute, only 36% of companies analyze traffic) [4, 5].

IDS can analyze and warn about what it can see, but if traffic is transmitted in encrypted form, the IDS can perform only limited verification based on a packet headers. The difficulty of considering the payload of a packet makes encrypted traffic one of the hardest problems for IDS.

Developers of leading companies and scientists are actively working on developing methods for detecting malicious activity in encrypted traffic. Cisco has developed a technology called ETA (Encrypted Traffic Analytics), which allows, based on network telemetry received from network equipment and machine learning algorithms, to classify encrypted traffic, incidentally separating the net traffic from malicious traffic.

The goal of this work is development the approach to decrypt SSL/TLS traffic for detection hidden threats.

## II. Basic Mechanisms for Checking SSL/TLS Traffic and Developing of Leading Companies

There are two basic mechanisms to inspect SSL/TLS traffic, depending on which key information and certificate is available, and how the inspection device is deployed on the network [6].

**1. Certificate Re-Signing** is the method used to inspect SSL/TLS on external servers for which organization does not have access to server certificates and private keys.

**Certificate Re-sign based Inspection** is relies on a validator that has a trusted CA certificate that can be used to sign SSL server certificates that have been intercepted and changed. This verification method is used for outbound communication of enterprise and intended for an external SSL server. The CA certificate used by the SSL Visibility Appliance can be self-signed, generated on box, and then signed by the corporate CA that processes the Certificate Signing Request from the SSL/TLS device or is generated by the corporate CA and loaded onto the appliance.

**Self-Signed Server Certificates** – the SSL Visibility Appliance can use re-sign certificate to process self-

signed server certificates or can simply replace keys in a self-signed server certificate, in which case the CA in a SSL/TLS device is not used, and the server certificate remains self-signed.

**2. Known Server Key** is relies on an SSL Visibility Appliance that has a copy of private keys and server certificates, and is used for incoming messages that are part of an organization that is destined for an internal SSL server. Certificates and keys can be easily and safely imported into the SSL Visibility Appliance application via the management interface or by using the third-party Enterprise Key and Certificate Management (EKCM) solution such as the Venafi's Trust Protection Platform to provide keys and certificates in the SSL Visibility Appliance.

Cisco Encrypted Traffic Analytics retrieves four basic data elements: a byte allocation, a TLS-specific function, a sequence of packet lengths and times and a source data packet. The unique architecture of Cisco's integrated circuit architecture (ASIC) provides the ability to retrieve these data items without slowing down data network. [2]

Cisco Stealthwatch Enterprise uses proxy servers, endpoint telemetry, NetFlow, traffic segmentation, policy and access mechanisms and much more to establish basic normal behavior of hosts and users in an enterprise. Stealthwatch can correlate traffic with a global threats to automatically identify suspicious traffic, infected hosts, command and control communications.

Stealthwatch supports a global risk map: a very wide behavioral profile about servers on the Internet, identifying the servers associated with attacks that can be used as part of an attack in a future. This is not a blacklist, but a whole picture in terms of security. Stealthwatch analyzes new encrypted traffic data elements in advanced NetFlow, using machine learning and statistical modeling. The global risk map and the Encrypted Traffic Analytics data elements are strengthened using pre-defined security analytics. Instead of decrypting traffic, Stealthwatch uses machine learning algorithms to identify malicious patterns in encrypted traffic in order to identify threats and improve response to an incident.

### III. BASIC MECHANISMS OF DECRYPTION INFORMATION

The SANS Institute described four approaches to decrypting SSL/TLS connections: 1) performing a check on the server itself; 2) the proxy server of the terminals; 3) decryption by the IDS itself; 4) autonomous tool for decrypting the connection [8-9].

**1. Perform the inspection on the server itself**. The easiest way to verify encrypted traffic is to use host-based IDS (HIDS) on the server, where a traffic belonging to this server is decrypted. HIDS can monitor server actions and look for changes in databases, unusual behaviors, system files or any critical data. Installing HIDS can add an additional load, which can negatively affect performance, especially for a load server.

**2. SSL/TLS termination proxy (reverse-proxy).** A reverse proxy is a server that acts as an intermediary between backend servers and clients. It accepts customer requests and retrieves resources effectively hiding backend servers from customers. The reverse-proxy server can be configured to encrypt SSL/TLS, acting as an SSL/TLS termination proxy, which removes the load from decrypting SSL/TLS connections, sending unencrypted traffic to associated servers. However, using the SSL/TLS proxy server allows to use IDS inside an internal network of the servers.

**3. The IDS performs the decryption**. IDS provides the ability to perform decryption process with a private key. This can be a pre-processor or plug-in that supports decryption and normalization of traffic before moving to a discovery mechanism. Currently, there is no Snort preprocessor for decryption process, although it is theoretically possible to develop such a pre-processor or plug-in (Snort FAQ, n.d.). However, a decryption function is available in some IDS devices, such as Juniper IDP.

**4. Standalone tool performs the decryption.** The Viewssld tool was used to decrypt an SSL/TLS connection using RSA key exchange. Viewssld is a free open source tool that can decrypt SSL/TLS traffic for IDS. It works by listening to an interface of specific IP address, decrypting encrypted traffic using the private server key and providing the decrypted traffic to IDS listening port. It does not support Diffie Hellman key exchange; it only supports RSA key exchange.

Symantec uses the Encrypted Traffic Management solution to eliminate blind of encrypted traffic [4]. The key component of this SSL Visibility Appliance solutions set is a high-performance SSL verification, decryption and management tool, scaling to 9 Gbps of SSL decryption and capable of simultaneously transmitting the decrypted information by several security tools.

The SSL/TLS verification and decryption capabilities provided by SSL Visibility Appliance allow existing security tools (IDS/IPS - Intrusion Prevention Systems, Malware Analysis, Next Gen Firewalls (NGFW), forensics, security analytics platforms) to access open text in SSL/TLS traffic, thereby allowing a security device to efficiently perform its function, even with SSL/TLS encrypted traffic.

The Gigamon offers the Security Delivery Platform GigaSECURE, in which security operations are allowed to use an unique "decryption zone" architectural approach to solve SSL/TLS decryption problem [7]. In a "decryption zone", the SSL/TLS traffic is decrypted once and fed to several security and operational tools for further analysis and verification, thereby eliminating unnecessary and repetitive cycles of decryption and re-encryption in an infrastructure. With this SSL/TLS decryption solution, Gigamon provides ubiquitous network visibility to identify malicious threats and provide decrypted traffic to an appropriate security tools for immediate analysis and management.

The Ponemon Institute asked respondents to assess the likelihood of specific attacks and the ability to resolve these attacks, which are shown in Table 1 [4].

TABLE I. THE LIKELIHOOD OF THE ATTACK AND THE ABILITY TO REMEDIATE THE ATTACKS

|  | The likelihood | |
|---|---|---|
|  | of attack | to resolve attack |
| S1. The attacker makes phishing threats more legitimate, and even informed recipients believe that using SSL makes it secure. However, by clicking on the link, it is transferred to an SSL server that is downloaded by malware that infects the client, because the malware traffic is encrypted and is not recognized by IPS. | 79% | 17% |
| S2. The attacker sends an encrypted stream of sensitive, secure and other critical data coming through the firewall over standart ports, for example 443 or 80, which the firewall is configured to accept, as they are approved ports. | 78% | 30% |
| S3. A number of malware families use encryption to hide network information, including sensitive data and passwords that they send to SSL servers. Encryption blinded the monitoring systems for these internal network activities. | 74% | 16% |
| S4. An attacker would confuse malicious communications when a worm, virus or botnet "home phones" sends stolen data to the host computer or downloads more malicious code or instructions. | 66% | 26% |
| S5. Using cross-site scripting, attackers steal cookies that can be used for a number of things, including session or account hijacking, cookie poisoning, changing user settings, and/or false advertising. All of this can be done by hiding within SSL-encrypted traffic. | 62% | 19% |

From Table 1 it can be seen that the resolution of the attack is rather small, compared with the probability of its occurrence.

The traffic decryption systems will continue to be more and more in demand, given the increased use of SSL/TLS encryption protocol. There are many situations when IT administrators should use packet validation, such as Wireshark. However, when application data is encrypted, troubleshooting application data becomes more complex.

## IV. PROPOSED APPROACH OF TLS TRAFFIC DECRYPTION

The approach described in this article is a approach for decrypting TLS 1.2 (1.3) traffic , which assumes that insider has access to a computer or network, or that an insider could put a bookmark on the victim's computer that can collect session data. Such conditions are needed to generate a session keys file, which will be used along with the corresponding intercepted traffic [10]. It is possible to intercept the victim's traffic being in any network part on a gap between server and a target of an attack.

Below is a description of the proposed approach implementation for decrypting TLS traffic. The implementation used the Wireshark traffic analyzer, which helps to analyze network performance, to diagnose problems, and has many other useful features.

To perform the experiment, a local network consisting of three subnets, a web server, an access server was taken. To connect computers from subnets to a web server a HTTPS by using TLS connection was used. In one of the subnets an additional access point was connected and a bookmark was made on one of the computers by sending it by e-mail. Then, using the Wireshark, the data in the network was read by additional point.

Earlier, by using traffic analyzers to decrypt TLS-traffic, several problems were encountered. One of these problems was the inability to easily analyze encrypted traffic like TLS. Also, to decrypt earlier versions of the TLS protocol 1.1, it was possible to point to Wireshark the private keys, if they were evaluable, and decrypt the traffic online, but this worked if only the RSA algorithm was used. The second problem is that the private key began to be received using the Diffie-Hellman protocol, which has high cryptographic strength, besides the private key can not be directly downloaded from the client, server or HSM (Hardware Security Module) where it is located. Because of such problems, it had to resort to actions associated with decrypting traffic through a man-in-the-middle attack (for example, through the sslstrip utility), which is of doubtful expediency.

Now to help get the keys comes logging session keys, which are used to encrypt and decrypt traffic. Getting such logs is not time consuming. Its can be obtained from browsers, for example, since recent versions Firefox and Chrome have learned to output data sufficient for derivation (reception) of session keys by which encrypts transmitted/received traffic, since symmetric encryption is used inside TLS. Strictly speaking, this is done not by browsers themselves, but by the NSS library in their composition; it specifies the format of the recorded files, as it described in [11]. In the case of traffic from java applications, you can create the required log files by examining the JVM (Java Virtual Machines) debug entries, namely by using the javax.net.debug option. After analyzing the debug entries received from this option, we can compile an NSS file with the appropriate format [11]. This format can be read and used by Wireshark to decrypt TLS records encrypted with the corresponding keys. To decrypt of TLS traffic we must have a file with logged session key records in NSS-format and a traffic analyzer, in this case Wireshark version 1.6.0 or higher.

Wireshark is very sensitive to the NSS file format, so it's better to carefully check whether the number of bytes in each line element converges, and if there are not any extra spaces anywhere, this can save time.

As for the traffic analyzer, we need to capture traffic after the keys are written to the log file, otherwise we will not be able to have the session keys that correspond to the captured TLS records. Also it is necessary to remember that the keys are "ephemeral". This means that they are suitable for only one TLS session, that is, the log from one session is not suitable for decrypting the traffic of another session. It is also necessary to use skills in working with Wireshark: tracking traffic exchange with a specific host and filtering by the necessary protocol to initially discard unnecessary packets passing through the listening interface. After it was possible to generate a file with session keys, we need to bind the file to Wireshark.

Now a package contents appeared tab "Decrypted SSL Data". Now, if we go to this tab we can see the text of the request. In addition, we can now select any package with the SSL/TLS protocol and select the function "Follow SSL Stream" in its context menu. The result is the contents of the packages in the form shown in Figure 1.

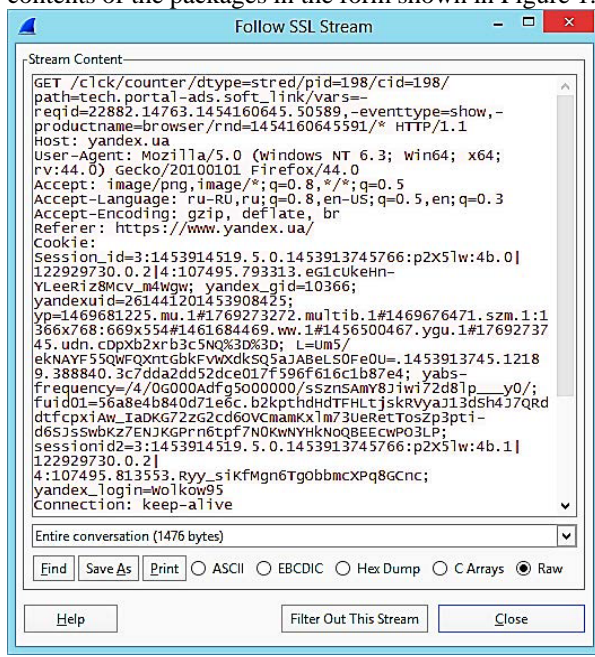

Figure 1.  View of the packages contents in Wireshark

As seen, in spite of the fact that communication passes through SSL/TLS we see the transmitted traffic and can export it for further analysis. The peculiarity of the described approach is that it is not necessary to intercept traffic on a computer that generates TLS traffic, it can be intercepted simply by being in a network and listening to it. A file with session keys can be obtained by putting a bookmark on the victim's computer or simply copying it, having access to the computer.

## Conclusion

SSL/TLS has become a universal standard for authenticating and encrypting messages between clients and servers. It is widely distributed in enterprises and is growing rapidly due to the rapid increase in mobile, cloud and web applications. However, SSL creates a security risk by introducing a "blind spot", which increases the risk of malware and unauthorized activity entering the organization. SSL / TLS verification is an important and desirable function for security analysts, but it has its costs.

To decrypt traffic, it is desirable to choose a way of decrypting traffic based on the needs and design of the network: on the server itself, the SSL/TLS termination proxy server or using the standalone tool or features added to the IDS. If the HIDS is installed on the server itself, it can add additional load, which can negatively affect performance, especially for a load server.

In this work, a approach for decrypting SSL/TLS traffic was developed, which can be used even in remote network listening. This apprioach was automated and allows you to decrypt data almost online.

In [12], an analysis of the TLS usage by malicious and corporate applications in the course of which millions of encrypted TLS streams were taken and a targeted study of 18 types of malicious programs consisting of thousands of unique samples of malicious programs and ten thousand malicious TLS streams. It is concluded that the use of TLS by malicious programs differs from benign use in enterprise settings and that these differences are effectively used in rules and classifiers of machine learning. In our future work, we plan to analyze encrypted SSL/TLS traffic using data science methods to detect unauthorized activity.